\documentclass[twocolumn,showpacs,amsmath,amssymb,floatfix]{revtex4}

\usepackage{graphicx}
\usepackage{color}

\newcommand{\lam}{\underline{\pmb \lambda}}

\newcommand{\expect}[1]{\left\langle #1 \right\rangle}
\newcommand{\Tr}{\ensuremath{\mbox{Tr}}}

\begin{document}

\title{Systematic errors in Gaussian Quantum Monte Carlo and a systematic study of the symmetry
    projection method}
\author{Philippe Corboz}
\affiliation{Institut f{\"u}r theoretische Physik, ETH Z\"urich, CH-8093 Z{\"u}rich, Switzerland}
\author{Adrian Kleine}
\affiliation{Institut f\"ur theoretische Physik C, RWTH Aachen, D-52056 Aachen, Germany}
\author{F. F. Assaad}
\affiliation{Institut f\"ur theoretische Physik und Astrophysik, Universit\"at W\"urzburg, Am~Hubland~D-97074~W\"urzburg}
\author{Ian P. McCulloch}
\affiliation{Institut f\"ur theoretische Physik C, RWTH Aachen, D-52056 Aachen, Germany}
\author{Ulrich Schollw\"ock}
\affiliation{Institut f\"ur theoretische Physik C, RWTH Aachen, D-52056 Aachen, Germany}
\author{Matthias Troyer}
\affiliation{Institut f{\"u}r theoretische Physik, ETH Z\"urich, CH-8093 Z{\"u}rich, Switzerland}

\begin{abstract}
Gaussian Quantum Monte Carlo (GQMC) is a stochastic phase space method for fermions with positive weights. In the example of the Hubbard model close to half filling it fails to reproduce all the symmetries of the ground state leading to systematic errors at low temperatures. In a previous work [Phys. Rev. B {\bf 72}, 224518 (2005)] we proposed to restore the broken symmetries by projecting the density matrix obtained from the simulation onto the ground state symmetry sector. 
For ground state properties, the accuracy of this method depends on a {\it large  overlap}  between the GQMC and exact density matrices.
Thus, the method is not rigorously exact.
We present the limits of the approach by a systematic study of the method for 2 and 3 leg Hubbard ladders for different fillings and on-site repulsion strengths. We show several indications that the systematic errors stem from non-vanishing boundary terms in the partial integration step in the derivation of the Fokker-Planck equation. Checking for spiking trajectories and slow decaying probability distributions provides an important test of the reliability of the results. Possible solutions to avoid boundary terms are discussed. 
Furthermore we compare results obtained from two different sampling methods:  Reconfiguration of walkers and the Metropolis algorithm.
\end{abstract}

\maketitle

\section{Introduction}
\label{intr}
One of the biggest unresolved problems in computational physics is the negative sign problem for fermionic and frustrated systems. Although it is not possible to solve it in general \cite{Troyer05} there is hope to find solutions for specific models. Gaussian Quantum Monte Carlo (GQMC) \cite{Corney04, corney_large} claims to be a sign-free ab-initio method for the general electronic structure problem. First results for the Hubbard model looked very promising. However, in a previous paper \cite{Assaad05} we have shown that in the vicinity of half filling systematic errors in the energy and other quantities occur, and that the method fails to reproduce the symmetries (SU(2) spin, translation, and lattice symmetries) of the Hamiltonian. This broken symmetry can be restored by projecting the low temperature density matrix from the simulation onto the ground state symmetry sector, such that excitations from other sectors are projected out. We have found that observables evaluated with the projected density matrix agree with exact ground state results, as we have shown for Hubbard models up to system sizes of 6x6.

One important aim of this paper is to investigate the origin of the symmetry breaking and the systematic errors. Our results suggest that they appear due to non-vanishing boundary terms in the partial integration step in the derivation of the Fokker-Planck equation. We show that changing the sampling method does not help to avoid this problem. 
Due to the inherent inaccuracy  of the density matrix as produced by the GQMC, it is not clear that symmetry projection schemes will produce correct ground state properties. 
It is therefore important to analyze the limits of this method. To this end we present a systematic study of Hubbard ladders.
 
The paper is organized as follows: The next section provides a summary of the method and the derivation of the stochastic differential equations (SDEs). A more detailed description was previously given in \cite{Assaad06}. 
Section \ref{systerr} addresses the origin of the systematic errors. We show that slow decaying power law tails in probability distributions can cause two different kinds of problems. First, they may lead to diverging variances of observables, making a Monte Carlo sampling useless. Second, boundary terms may appear in the partial integration step in the derivation of the Fokker-Planck equation. In the presence of boundary terms the SDEs are no longer valid, and by neglecting them a systematic error is introduced. This problem has been encountered before in the context of stochastic phase space methods for bosonic systems \cite{Gilchrist97} and was solved for specific models with the help of stochastic gauges \cite{Deuar02}. A side effect of boundary terms is the presence of spiking trajectories, therefore checking for spikes in the phase space variables is an important test of the reliability of the results.
In section \ref{metropolis}  we discuss results from the Metropolis algorithm which leads to the same systematic errors as the reconfiguration scheme of walkers \cite{sorella} we usually use.

Empirically we have seen that one of the major consequences of  fat tailed distributions shows up in the violations of symmetries.   Hence,  imposing symmetry projections prior to measurements can potentially correct for sources of systematic errors. 
In section \ref{hubbardladders} we present a systematic study of the GQMC method with symmetry projection (PGQMC) for 2 and 3 leg Hubbard ladders for different fillings and on-site repulsion strengths, and compare the results with Density Matrix Renormalization Group \cite{White92, Schollwoeck05} (DMRG) calculations. Symmetry projection is successful in removing systematic errors in all cases where the overlap of the density matrix with the ground state symmetry sector is not too small. However, for a small overlap systematic errors may still be present. In the outlook in section \ref{outlook} we refer to recent promising improvements of the projection method by Aimi and Imada~\cite{Aimi07}.

\section{Summary of the method}
\label{method}
Let us briefly recall the derivation of the SDEs which is a standard procedure for various stochastic phase space methods. A more detailed derivation can be found in Ref. \cite{corney_large, Assaad06}. The starting point is an expansion of the system density
operator in an over-complete operator basis 
\begin{eqnarray}
\label{Den_matrix}
 \hat{\rho}(\tau) & = & 
 \int d \lam P(\lam,\tau) 
 \hat{\Lambda}(\lam),
 \end{eqnarray}
where $\tau$ is the inverse temperature and the probability density $P$ is normalized $ \int {\rm d} \underline{\pmb \lambda} 
P(\underline{\pmb \lambda},\tau) = 1$. The $\hat \Lambda(\lam)$ are the Gaussian operator basis elements of the normal ordered form 
\begin{equation}
\hat{\Lambda}( {\bf n},\Omega )  = \Omega \det( {\bf 1} - {\bf n}  ) :e^{ - \hat{{\bf c}}^{\dagger} \left(  {\bf 2} + \left( {\bf n}^T - {\bf 1} \right)^{-1} \right) \hat{{\bf c}}   } :, 
\end{equation}
with $\hat{{\bf c}}^{\dagger}$ (respectively $ \hat{{\bf c}}$) being an $N_s$ dimensional vector of creation (respectively annihilation) operators, ${\bf n}$ is an $N_s \times N_s$ real matrix of phase space variables and $N_s$ denotes the number of states. $ \det( {\bf 1} - {\bf n}  )$ is the normalization term such that $\text{Tr}[\hat{\Lambda}({\bf n},\Omega)]=\Omega$. Thus, $\Omega$ plays the role of a weighting factor.

The imaginary time evolution of the density operator reads:
\begin{equation}
        \frac {d }{d \tau}  \hat{\rho}(\tau)  =          
- \frac{1}{2} \left[\hat{H}, \hat{\rho}(\tau)  \right]_{+}.
\label{fp1}
\end{equation}
Introducing the expansion \eqref{Den_matrix} for $\hat \rho$ leads to
\begin{equation}
\frac {d }{d \tau} \int{d} \lam  P(\lam,\tau) \hat \Lambda(\lam) =
-\frac{1}{2}\int d
\lam  P(\lam,\tau)  \left[\hat{H},  \hat \Lambda(\lam) \right]_{+}.
\label{fp2}
\end{equation}
With the help of differential properties of the operator basis derived in Ref. \cite{Corney04} the action 
of the Hamiltonian on the operator basis element can be transformed into an operator $L$ containing first and second order derivatives with respect to the phase space variables $\lam$, and we formally write
\begin{equation}
\int{d} \lam  \frac {d }{d \tau} P(\lam,\tau) \hat \Lambda(\lam) = \int{d}
\lam  P(\lam,\tau) L[\hat \Lambda(\lam)].
\label{fp3}
\end{equation}
In the next step we perform a partial integration where we take all derivatives acting on the basis element in front of P, which is denoted by the new operator $L'$:
\begin{equation}
\int{d}\lam  L'[P(\lam,\tau)] \hat \Lambda(\lam) + 
\text{boundary terms.}
\label{fp4}
\end{equation}
Depending on the nature of the distribution $P$, boundary terms from the partial integration arise. Let us first assume that these boundary terms vanish, so that we can compare integrands on both sides to obtain
\begin{equation}
\frac{d}{d\tau}P(\lam,\tau) = L'[P(\lam, \tau)],
\label{fp5}
\end{equation}
where we have omitted the Gaussian basis element on both sides.  
This corresponds to a Fokker-Planck equation describing the evolution of the distribution
function $P$ in (imaginary) time. If $L'$ is of the form
\begin{equation}
L' = - \sum_\alpha \frac{\partial}{\partial \lambda_\alpha} A_\alpha +
\frac{1}{2}\sum_{\alpha\beta k}
\frac{\partial}{\partial \lambda_\alpha} B_\alpha^k 
\frac{\partial}{\partial \lambda_\beta} B_\beta^k
\label{L'}
\end{equation}
with $A_\alpha$ and $B_\alpha$ real functions we can derive real
valued (Stratonovich) SDEs 
\begin{equation}
\label{sde.eqn}
d\lambda_\alpha(\tau) = A_\alpha(\lam)d\tau + \sum_k B_\alpha^k(\lam)dW_k(\tau)
\end{equation}
with Wiener increments $dW_k(\tau)$ defined by the correlations $\langle
dW_k dW_{k'}\rangle = d\tau \delta_{k k'}$
and the mean $\langle dW_k(\tau)\rangle=0$. 
The explicit forms of the functions $A_\alpha(\lam)$ and $B_\alpha^k(\lam)$ for the Hubbard model can be found in appendix \ref{app1}.
The form of $L'$ is not unique but can be
modified by gauge degrees of freedom. In Ref. \cite{Corney04} the ``Fermi gauge" $\hat n_{ii\sigma}^2-\hat n_{ii\sigma}=0$
is used to obtain real valued SDEs with positive weights. Adding such terms clearly does not
modify the expectation value of the Hamiltonian, but changes the resulting
Fokker-Planck equation.

\section{Sources of systematic errors}
\label{systerr}
\subsection{Systematic errors in the Hubbard model}
We have tested the GQMC method for the Hubbard model given by the Hamiltonian
\begin{equation}
\hat H = - t \sum_{\langle i,j \rangle, \sigma} \hat n_{ij\sigma} + U \sum_i \hat n_{ii\uparrow}
\hat n_{ii\downarrow} - \mu \sum_{i,\sigma} \hat n_{ii\sigma},
\end{equation}
with nearest neighbor hopping strength $t$, on-site repulsion $U$ and chemical potential $\mu$. 
The corresponding stochastic differential equations derived under the assumption of vanishing boundary terms can be found in appendix \ref{app1}.

\begin{figure}[h]
  \includegraphics[height=0.27\textheight]{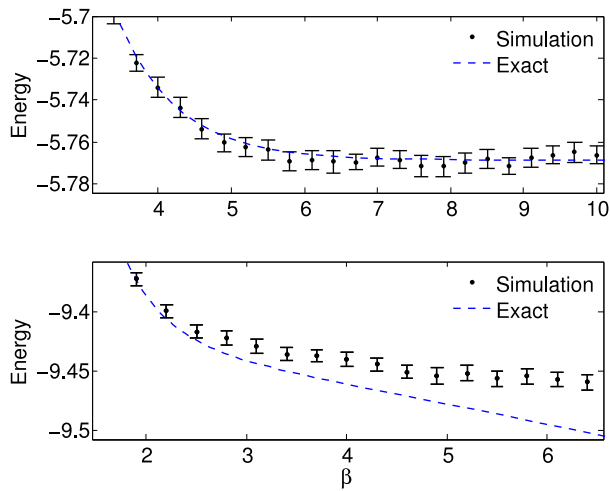}
  \caption{(Color online) Energy as a function of inverse temperature $\beta$ obtained from exact
    diagonalization (dashed lines) and GQMC runs (solid lines) for a $2\times 2$ Hubbard model. In the upper plot the system is far away from half filling: $U=1$, $t=1$, $\mu=-1$, averaged over 40,000 trajectories. The exact result is reproduced within the statistical error bars. In the lower plot, close to half filling, the energy from the simulation is systematically too high ($U=4$, $t=1$, $\mu=1$, averaged over 480,000 trajectories).}
\label{energy.fig}
\end{figure}

As already pointed out in Ref. \cite{Assaad05} the GQMC method works well for small electron interaction $U/t$ and away from half filling (Fig.~\ref{energy.fig}, upper plot). In this regime the ground state is very well described by a paramagnetic mean field solution which is exactly reproduced by the GQMC approach. Close to half filling and with a big on-site repulsion  the simulation results exhibit systematic errors (Fig.~\ref{energy.fig}, lower plot). 
The energy agrees with the exact result down to a certain temperature, but for lower temperatures the mean energy is systematically too high. In Ref. \cite{Assaad05} it was found that the solution of the simulation does not preserve SU(2) spin rotation symmetry. This gave the motivation to develop the projection scheme as described in section \ref{hubbardladders}. But the reason for this symmetry breaking has not been found so far. In section \ref{sec:BT} we suggest that the systematic errors and the symmetry breaking stem from non-vanishing boundary terms.

\subsection{Power law tails in the probability distribution of observables}
In this section we discuss how a power law tail in the probability distribution of an observable X can lead to problems in the Monte Carlo sampling. The error on the expectation value $\langle X \rangle$ obtained from a Monte Carlo simulation is given by $\Delta X/\sqrt{M}$ where M is the number of independent samples and $\Delta X$ the variance
\begin{equation}
\Delta X = \left( \langle X^2 \rangle- \langle X\rangle^2\right)^{1/2}.
\end{equation} 
If the variance of X diverges, then also the error bar of our Monte Carlo result diverges. 
Thus, to obtain a meaningful result from the sampling, the variance has to be well defined. The m$^{\text{th}}$ moment of the probability distribution $P(X)$ is given by  
\begin{equation}
\langle X^m \rangle= \int X^m P(X)dX.
\end{equation} 
If the probability distribution exhibits a power law tail $P(X) \propto X^{-p}$ for $X \rightarrow \infty$ then  only moments $m<p-1$ of $P(X)$ converge, because
\begin{eqnarray}
\int X^m P(X)dX &\rightarrow& \int X^m X^{-p} dX  = \nonumber \\ 
=\int X^{(m-p)}dX  &\rightarrow& \infty, \quad \text{for } m-p \ge -1. 
\end{eqnarray} 
Therefore, to obtain a finite mean corresponding to the first moment ($m=1$) of $P(X)$,  $p$ has to be bigger than $2$. For a finite variance the integral has to converge also for $m=2$, which inquires $p>3$. If $p<3$ we do not obtain a meaningful result from a Monte Carlo sampling.
 
\begin{figure}[t]
\centering
\includegraphics[width=0.47\textwidth]{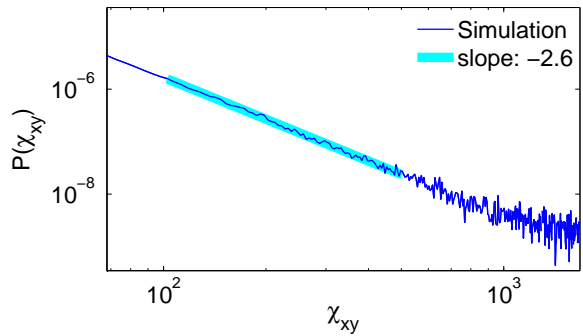}
\caption{(Color online) Log-log plot of the distribution of the transverse spin susceptibility for the $2\times 2$ Hubbard model at half filling with $U/t=4$ at low temperatures ($\beta=10$). The slope of the power law tail yields $p \approx 2.6$, therefore the variance is not defined. }
\label{chi_xy_hist}
\end{figure} 

We have found a diverging variance for the transverse spin susceptibility $\chi_{xy}$ at low temperatures in a simulation exhibiting systematic errors. The slope of the distribution $P(\chi_{xy})$ in the log-log plot in Fig.~\ref{chi_xy_hist} yields $p \approx 2.6$ such that the second moment is not defined. 
We also observe systematic errors in the energy. However, the variance of the energy is always well defined in our simulations. Thus, the systematic errors found in the energy cannot be explained by an ill defined Monte Carlo sampling. In the next section we demonstrate that power law tails in the distribution of the phase space variables can cause a different kind of problem, namely non-vanishing boundary terms.

\subsection{Non-vanishing boundary terms}
\label{sec:BT}
Boundary terms (BTs) from the partial integration in Eq. \ref{fp4} appear if certain moments of the high-dimensional probability distribution $P(\lam, \tau)$ do not converge. Initially, at infinite temperature,  $P(\lam,\tau=0)$ is a delta-function, for which all moments converge and therefore there are no BTs at the beginning. As the distribution $P$ evolves in imaginary time it becomes broader and we show below, that it develops slow decaying power law tails, such that BTs cannot be excluded anymore. Typically BTs appear after a specific imaginary time $\tau>\beta_{BT}$. But other situations are known \cite{Gilchrist97}, where the BTs are present only for a short time and disappear again. What are the implications of BTs on our simulation? We have derived SDEs under the assumption of non-vanishing BTs. As soon as they become non-negligible, the SDEs are strictly speaking no longer valid, such that we obtain a wrong distribution $P(\lam, \tau)$ for $\tau>\beta_{BT}$. We suggest that this is the origin of the systematic errors in the energy and in other quantities. The problem of boundary terms has been discussed in detail for several bosonic systems \cite{Gilchrist97, Deuar02}. The following analysis is done in a similar spirit as for these systems.

\begin{figure}[t]
\centering
\includegraphics[width=0.45\textwidth]{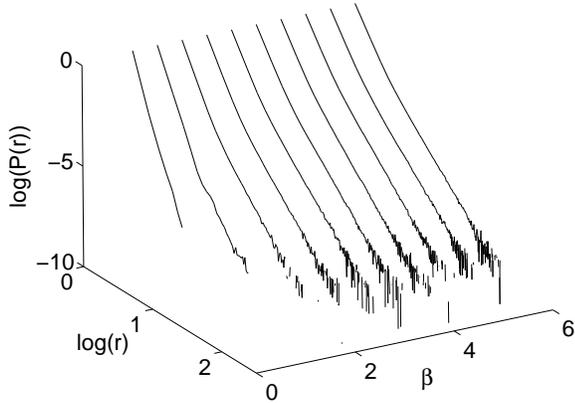}
\caption{Appearance of a power law tail in the distribution $P(r)$ (log-log plot) at low temperatures (large $\beta$). The simulation parameters are $U=4$, $t=1$, $\mu=1$ and 480,000 trajectories.}
\label{Pr}
\end{figure}

\begin{figure}[h]
\centering
\includegraphics[width=0.47\textwidth]{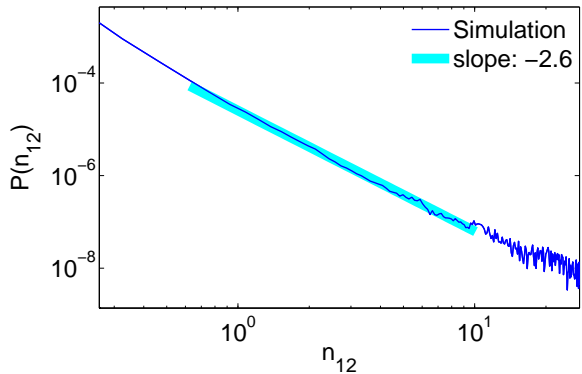}
\caption{(Color online) Power law tail in the distribution $P(n_{12\uparrow})=P(n_{12\downarrow})$ for 2 site system at half filling for $U/t=100$, $\beta=3$, 800,000 walkers. }
\label{tail_2site.fig}
\end{figure}

A first test for the presence of BTs is to measure the radial averaged distribution $P(r) \sim r^{-p}$ with $r=\sqrt{\sum_{ij\sigma} n_{ij\sigma}^2}$ at different inverse temperatures $\beta$. The power law tail in Fig. \ref{Pr} reaches a slope of $p \approx 3.8$. This already indicates that power law tails are also present in the high-dimensional distribution $P(\lam)$ of all phase space variables. 
From the partial integration step in Eq. \ref{fp4} we can find the explicit expressions for the BTs. They are  of the form
 \begin{equation}
 \label{formalbt}
 \left. \int d\lam_{(\alpha)} M_{(\alpha)}(\lam^{(\alpha)}) P(\lam) \hat \Lambda(\lam)  \right|_{\text{Boundary}}
\end{equation}
where $\alpha$ enumerates the different BTs, $M_{(\alpha)}$ depends on the phase space variables up to fourth order and the integral is taken over all phase space variables except the one the partial integration has been carried out for. The basis element $\hat \Lambda(\lam)$ also depends on the phase space variables. It includes terms in $n_{ij\sigma}$ up to $2N_s$ th order.
We present here one specific example of a non-vanishing boundary term for a 2 site model at half filling with $U/t=100$: 
 \begin{widetext}
\begin{equation}
\lim_{a_{11 \uparrow}\rightarrow \infty} \lim_{a_{12 \uparrow}\rightarrow \infty} 
... \lim_{a_{NN \downarrow}\rightarrow \infty} \left. \int_{-a_{11 \uparrow}}^{a_{11\uparrow}}dn_{11\uparrow }\int_{-a_{13\uparrow}}^{a_{13\uparrow}}dn_{13\uparrow}\int_{-a_{14\uparrow}}^{a_{14\uparrow}}dn_{14\uparrow}
...\int_{-a_{NN\downarrow}}^{a_{NN\downarrow}} dn_{NN\downarrow} \hspace{0.1cm}
n_{12\uparrow}^3 P(\lam)  \right|_{-a_{12\uparrow}}^{+a_{12\uparrow}}.
\label{explicitbt}
\end{equation}
\end{widetext}
Comparing with Eq. \ref{formalbt} this term corresponds to $M_{(\alpha)}(\lam^{(\alpha)})=n_{12 \uparrow}n_{12 \uparrow}$, and from the expansion of  $\hat \Lambda(\lam)$ we took the term proportional to $n_{12 \uparrow}$, leading to an integrand which is cubic in the phase space variable $n_{12 \uparrow}$. 
Eq. \ref{explicitbt} stems from a partial integration with respect to the variable $n_{12\uparrow}$. The remaining integral is taken over all variables $n_{ij \sigma} \neq n_{12 \uparrow}$, yielding the marginal distribution $P(n_{12 \uparrow})$:
\begin{equation}
\label{bt2}
\lim_{a_{12\uparrow}\rightarrow \infty}\left. n_{12 \uparrow}^3 P(n_{12 \uparrow})  \right|_{-a_{12 \uparrow}}^{+a_{12 \uparrow}}
\end{equation}
The fit to the power law tail of the distribution $P(n_{12 \uparrow})$ in Fig. \ref{tail_2site.fig} yields an exponent of $p\approx 2.6$. Therefore, with $P(n_{12 \uparrow}) \sim n_{12 \uparrow}^{-2.6}$ this term does not vanish and we therefore cannot neglect it in the partial integration step. 
This BT is particularly simple to analyze because it involves only a one-dimensional distribution. For other BTs with mixed variables one would need to study a distribution of several variables. 
By studying all possible terms appearing one could find the minimal exponent $p_{BT}$ necessary to exclude BTs. But as discussed in Ref. \cite{Gilchrist97} there are simpler ways to detect BTs as we show in the next section.

The question may arise why power laws actually occur. The problem lies in the strong multiplicative noise term in our SDEs. The noise is amplified by the phase space variables $n_{ij\sigma}$ themselves (the diffusion term $B_\alpha^z$ in Eq. \ref{sde.eqn} is even quadratic in $n_{ij\sigma}$). This has strong consequences on the functional form of the distribution $P(\lam)$. In Ref. \cite{Biro05} it is shown that the smallest multiplicative noise in the linear Langevin equation changes the Gaussian stationary distribution to one with a power-law tail. Power laws can arise from multiplicative stochastic processes as naturally as Gaussian distributions from processes with additive noise \cite{Levy96}. 
The question remains if at finite time $\tau$ the distribution will always have a finite cut-off at a certain distance such that the BTs would always vanish. However, we have not found such a cut-off in our simulations. 

\subsection{Spiking trajectories}
\begin{figure}[h]
  \includegraphics[height=0.25\textheight]{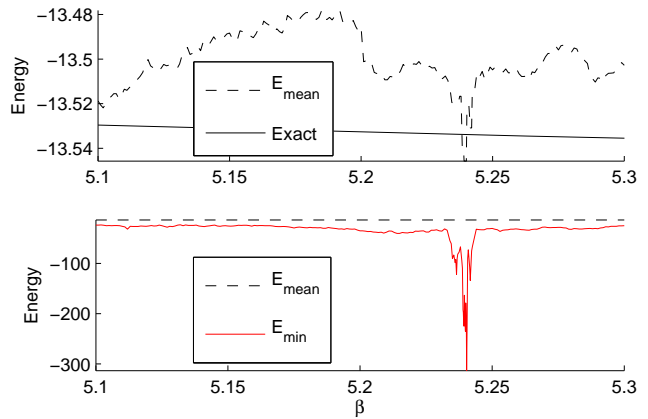}
  \caption{(Color online) Upper figure: Mean energy from the simulation compared with the exact energy. The spike at $\beta \approx 5.24$ is caused by a single extreme trajectory shown in the lower figure (solid line). The spike occurs within a few time steps. The simulation is done at half filling for a $2\times 2$ lattice with $U/t=4$, and 10,000 trajectories.}
\label{spike.fig}
\end{figure}

There are other indications that the boundary terms in equation \eqref{fp4} do not vanish. 
According to Ref. \cite{Gilchrist97} spiking behavior of the trajectories imply that BTs are likely to be significant. Such near-singular trajectories  do large excursions in phase space for a very short time (within a few time steps). 
Spikes could also stem from an unstable integration scheme. It is therefore important to use a stable integrator. Fig.~\ref{spike.fig} shows an example of a sharp spike in the energy. Such extreme trajectories lead also to a sudden increase of the statistical error of observables.

As mentioned in the last section, there are no BTs at the beginning of the simulation, as we start from a delta function for the distribution $P(\lam, \tau=0)$. They appear at a specific inverse temperature $\beta_{BT}$ as we integrate the SDEs towards lower temperatures. The first appearance of a spike should give an estimate of $\beta_{BT}$. In the example of Fig.~\ref{energy.fig} (lower plot) the first spiking walker shows up for $\beta_{BT} \approx 1.5$, which is in good agreement with the inverse temperature, at which the energy starts to deviate from the exact result.

For small interaction $U/t$, where we obtain correct results, not a single spike can be observed. Therefore testing for spikes provides a good indicator, whether the GQMC results are reliable. 

\subsection{Stochastic gauges}
As already mentioned for a given Hamiltonian the SDEs are not unique, but different choices of "gauges" are possible. Thanks to the overcompletness of the basis several solutions of the distribution $P(\lam,\tau)$ exist. With appropriate gauges the boundary terms could possibly be removed \cite{Corney05}, leading to a more compact distribution. The so-called drift gauges could be used to avoid nearly-diverging trajectories which cause power law tails. The tradeoff is to introduce noise into the equation for the weight. Stochastic gauges have been successfully applied for several models \cite{Deuar02}. A future analysis will show if similar techniques can be applied to the Hubbard model to solve the present problems. 
\vspace{3cm}


\section{Metropolis algorithm}
\label{metropolis}
The equation for the weight $\Omega(\tau)$ (equation (\ref{sdes.eqn}) from the appendix) of a walker can be integrated. The weight as a function of time then reads 
\begin{equation}
\label{weight_int.eqn}
\Omega(\beta) = \exp \left( - \int_0^\beta H({\pmb n}(\tau)) ~ d\tau \right).
\end{equation}
The weight and the variance of the weight thus grow exponentially, yielding the need of some importance sampling. We usually use a
reconfiguration scheme similar to the one used in the Green function Monte Carlo method \cite{sorella}. 
A further way to sample the distribution is to use the Metropolis - Hasting algorithm which was recently proprosed by Dowling et al. \cite{dowling}.
In this section we want to briefly summarize the basic ideas of the algorithm within the framework of the GQMC and then present some results.

\subsection{Metropolis Algorithm}
Starting from an arbitary state $s_n$ one uses a suitable candidate generation function
$ q(s,s') $ in order to create a proposed step. This step is then accepted $ s_{n+1} = s'$ with the probability
\begin{equation}
\label{metropolis_hasting.eqn}
\alpha = \min \left( \frac{\pi(s') q(s',s)}{\pi(s) q(s,s')},~ 1 \right).
\end{equation}
If the move is rejected, the old state is kept $ s_{n+1} = s_n$. One can prove that this algorithm fulfills the detailed balance so the resulting
chain samples the target density $\pi$ correctly \cite{hasting}.

\begin{figure}[h]
\begin{center}
\includegraphics[width=.4\textwidth]{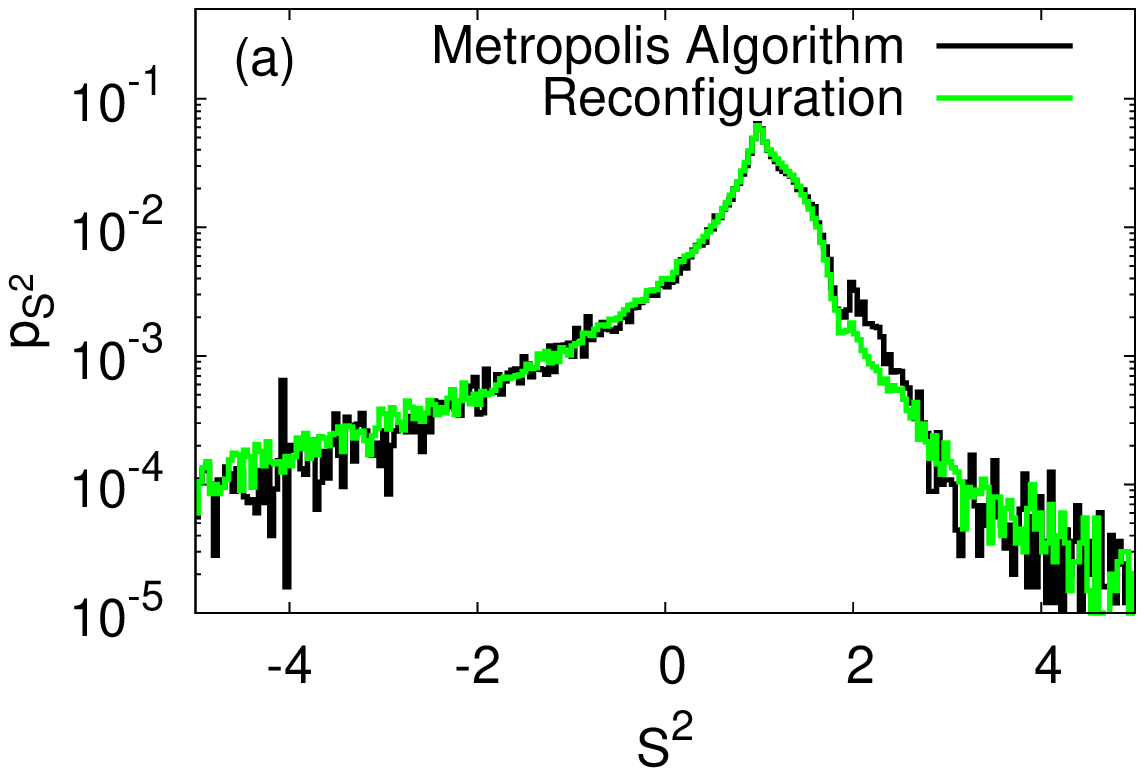}
\includegraphics[width=.4\textwidth]{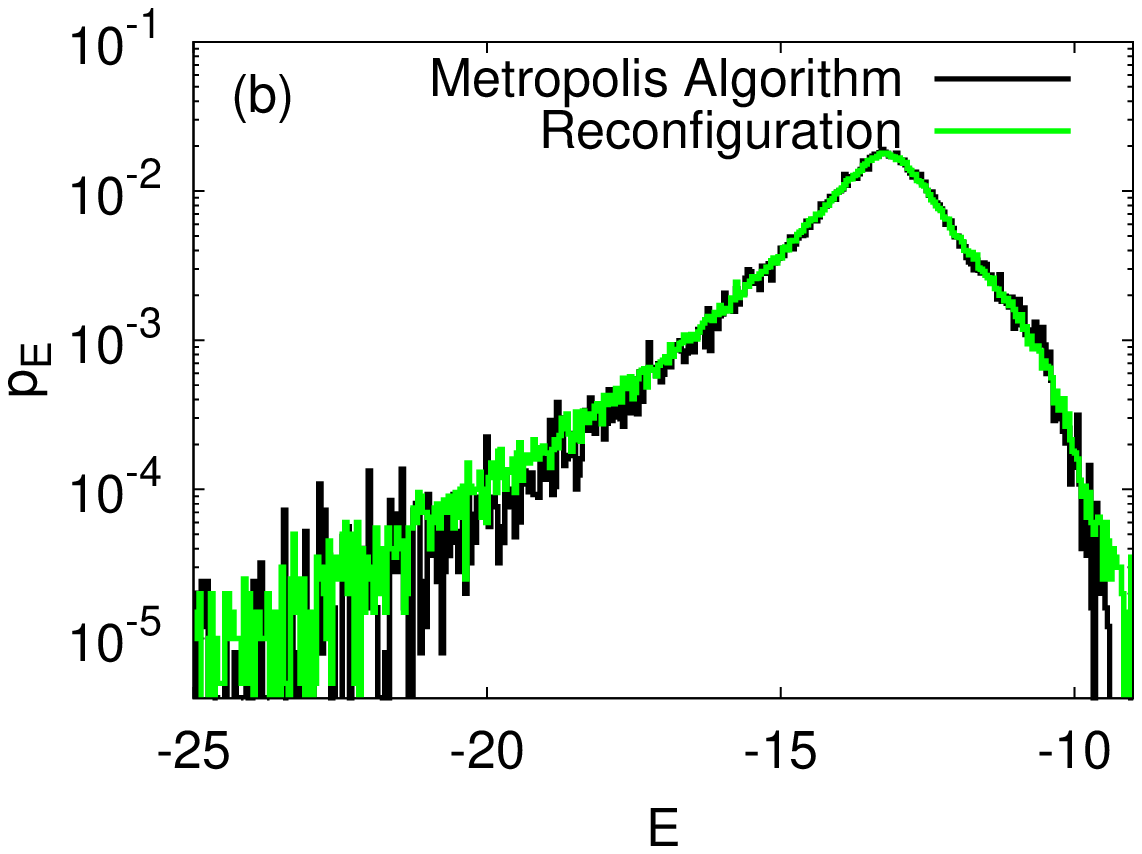}
\end{center}
\caption[]{(Color online) Probability distributions of the total spin squared $\hat {\boldsymbol S}^2$ (a) and the energy $E$ (b) evaluated at $\beta = 20$ obtained
by using the Metropolis algorithm and the reconfiguration scheme ($ 2 \times 2 $ Hubbard model, $U / t = 4$, half filled, 100,000 samples each)}

\label{metropolis_hist.fig}
\end{figure}

The usual way to solve SDEs (as in Eq. (\ref{sde.eqn})) is to
apply an Euler -- Marujama method (either implicit or explicit). For each time step one needs a fixed number of Gaussian distributed random numbers. Let this
number be $M$ and consider $N$ time steps. Then one sample is given by a noise vector $\vec{w} \in {\bf R}^{MN} $ and the phase space variables for the
final time $\tau = N ~\Delta\tau$ can be considered as a function of the noise vector, i.e. ${\bf n}(\vec w)$ and $\Omega(\vec w)$. $\vec w$ is a normal distributed
random variable with a probability density given by
\begin{equation}
P(\vec w) ~ \sim ~ \exp \left( - \frac{\vec{w}^2}{2} \right).
\end{equation}
The expectation value of an observable $\hat O$ can thus be written as
\begin{eqnarray}
\expect{\hat O} & = & \frac{\Tr \hat\rho ~ \hat O}{\Tr \hat\rho}
=\frac{\expect{\Omega(\vec w) ~ O(\vec w)}}{\expect{\Omega(\vec w)}} \nonumber\\
& = & \frac{\int P(\vec w) ~ \Omega(\vec w) ~ O(\vec w) ~ d^{MN}\vec w}{\int P(\vec w) ~ \Omega(\vec w) ~ d^{MN}\vec w}.
\end{eqnarray}
Now, one can apply the Metropolis-Algorithm to create samples $\left(\vec{w_i}\right)_i$ which are distributed like
\begin{equation}
\pi(\vec{w}) = P(\vec{w}) ~ \Omega(\vec{w}),
\end{equation}
and finally the expectation value of the observable is given by $ <O> = \frac{1}{N} \sum^N_{i=1} O(\vec{w_i}) $. It is convenient to use a candidate generating
density which obeys
\begin{equation}
\frac{q(\vec{w}, \vec{w}')}{q(\vec{w}', \vec{w})} = \frac{P(\vec{w}')}{P(\vec{w})}.
\end{equation}
The acceptance rate is then just the quotient of the weights $ \Omega$, i.e.
\begin{equation}
\alpha = \min \left( \frac{\Omega(\vec{w}')}{\Omega(\vec{w})},~ 1 \right).
\end{equation}
We used a simple candidate generation function which alters each component of the noise vector with a fixed probability $r$, i.e. drawing
approximately $ r M N$ new Gaussian numbers. One can easily adapt this algorithm to an adaptive step size (see appendix \ref{ats}) using a dynamical enlarged noise space.

\begin{table}[h]
\begin{tabular}{|c|c|c|c|}
\hline
  & Metropolis Sampling & Reconfiguration & Exact   \\
\hline
$E$ & $-13.57 \pm 0.01$ & $-13.53 \pm 0.01$ & $-13.615$ \\
$S^2$ & \phantom{a}$0.77$ & \phantom{a} $0.72$ & $0.295$ \\
\hline
\end{tabular}
\caption{Comparison between the Metropolis and the reconfiguration results, ($ 2 \times 2 $ Hubbard model at half filling, $U / t = 4$, evaluated at $ \beta = 20 $,
100,000 samples). The error bar on $S^2$ is ill defined because the variance of $S^2$ diverges.}
\label{table_metro.dat}
\end{table}

\subsection{Results}

Now we want to present some results using the Metropolis algorithm. The system which we discuss here is a $ 2 \times 2 $ Hubbard model at half filling with $U / t = 4$. The behaviour of this model is representative of that seen in other systems, and seems to be quite generic. The SDEs are solved using an implicit Euler scheme with an adaptive time step with $\Delta \tau_{\text{max}} = 5\cdot 10^{-4}$ (see appendix \ref{ats}) . 

The data using the reconfiguration scheme is obtained from 100,000 walkers and by applying the scheme every $\Delta \tau_{rc} = 0.05$. For the Metropolis algorithm
a typical chain length is 1,000 after a burn-in time of the order of 100 Metropolis steps. The Metropolis algorithm has one major drawback compared to the reconfiguration scheme, namely one needs to fix
a specific target time $\beta$. When using the reconfiguration, one is in principle able to obtain values for any intermediate time and not just for the final one. Therefore the computational effort to calculate the observables for all times $\beta = [0,~\beta_{final}]$ is much bigger for the Metropolis algorithm than for the reconfiguration. An interesting variant of the Metropolis algorithm which allows the calculation of observables for intermediate times is presented in Ref. \cite{Aimi07}.

Table \ref{table_metro.dat} shows the expectation values of the total spin squared $\hat{\bf S}^2$ and $\hat H$ at the target time $ \beta = 20 $.
The observable $\hat{\bf S}^2$ is chosen since after
some time the variance of this observable is not finite anymore, thus yielding a good test whether changing the sampling method improves the results. For the results for the Metropolis algorithm
100 Markov chains with each 1,000 steps after the burn-in time are used, thus yielding 100,000 samples, which is the same number as the one used for the reconfiguration scheme.
One clearly sees that the results do not change significantly, the energy is slightly improved but $\hat S^2$ gets worse. 
To further investigate the effect of the sampling method, the probability distributions of the two observables were also calculated (see Fig. \ref{metropolis_hist.fig}). Again both methods produces
almost the same probability distribution, especially the slow decaying power law tails of $\hat S^2$ are still present.
Using the Metropolis sampling therefore does not seem to change the results significantly.

\section{Simulation of Hubbard ladders}
\label{hubbardladders}
In Ref. \cite{Assaad05} we have found that GQMC fails to reproduce all the symmetries of the Hamiltonian at low temperatures. 
We proposed to restore the broken symmetries by projecting the density matrix obtained from the simulation $\hat \rho_{\text{sim}}$ onto the ground state symmetry sector,
\begin{equation}
\hat \rho_{\text{proj}}=\hat P \hat \rho_{\text{sim}} \hat P^\dagger,
\end{equation}
where $\hat P$ is the corresponding projection operator. 
For example projecting the density matrix onto the $S=0$ sector filters out spin excitations and restores the SU(2) rotation symmetry in spin space. 

The discussion in section \ref{systerr} suggests that the symmetry breaking is directly related to the presence of boundary terms. In Ref. \cite{Assaad05}  we obtained correct results for the ground state by an appropriate symmetry projection, which implies that even in the presence of boundary terms, the correct ground state is still included in the density matrix, but mixed with excited states, which we can project out. The question remains, if such a projection can always be done. The aim of the current section is to show the limits of this projection method (PGQMC) for the example of Hubbard ladders for various lengths, interaction strengths and doping. Energy and correlation functions are compared with calculations from the Density Matrix Renormalization Group (DMRG \cite{White92, Schollwoeck05}) method which provides high precision results for quasi one dimensional systems. The DMRG results are calculated in a matrix product state basis using both the SU(2) spin and the SU(2) pseudospin symmetry \cite{McCulloch02}.

The PGQMC simulations in this section are done with 8,000 - 32,000 walkers, an adaptive time step with $\Delta \tau_{\text{max}}=5 \cdot 10^{-4}$ (see appendix \ref{ats}) and an explicit Euler integration scheme. For some examples crosschecks have been made with more walkers, with an implicit Euler scheme and different quantization axis. The error bars stem from averaging over several projections at different imaginary times. They do not take into account the discretization error in the symmetry projection. For the comparison with the exact values we check if they are within two standard deviations ($2\sigma$). If they lie outside we have to assume that besides the statistical error, there is also a systematic error present. 

Projection onto the ground state is only possible if there is a finite overlap ($\Tr [{\hat  \rho_{\text{proj}}}]/\Tr [{\hat  \rho_{\text{sim}}}]$) between the density matrix and the ground state symmetry sector, or in other words, if the density matrix from the simulation contains a finite contribution of the ground state, which we can extract by the projection. The projection is always made onto the $S=0$ sector and onto all possible momentum and parity sectors. Note that we only have translational symmetry along the x-axis, whereas along the y-axis we can distinguish odd and even parity. The ground state sector is the one with lowest energy, and we found that this  sector always has the biggest overlap with the density matrix.

\subsection{Two leg ladders}
\subsubsection{L=4 and varying U}
\begin{figure}[t]
\includegraphics[width=0.475 \textwidth]{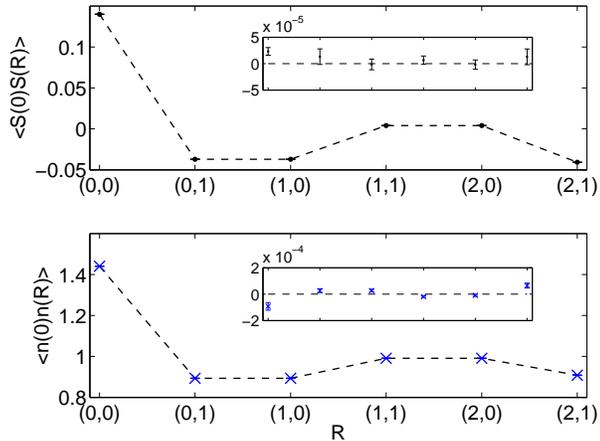} 
\caption{(Color online) Spin-spin and charge-charge correlation function of the half filled 4x2 Hubbard ladder for U/t=1 (without symmetry projection) showing perfect agreement with DMRG. The deviation from the DMRG result (dashed line) is shown in the inset.}
\label{per4_1.fig}
\end{figure}

As already pointed out GQMC works well in the weak interacting case (small $U/t$) and systematic errors appear for large $U/t$.  For $U/t=1$ the GQMC simulation results for the spin-spin and charge-charge correlation functions in Fig.~\ref{per4_1.fig} agree with the DMRG results even without symmetry projection. Also the energy $E_{GQMC}=-10.117 \pm 0.001$ is correct compared to $E_{DMRG}=-10.118$. 

\begin{figure}[b]
\includegraphics[width=0.475 \textwidth]{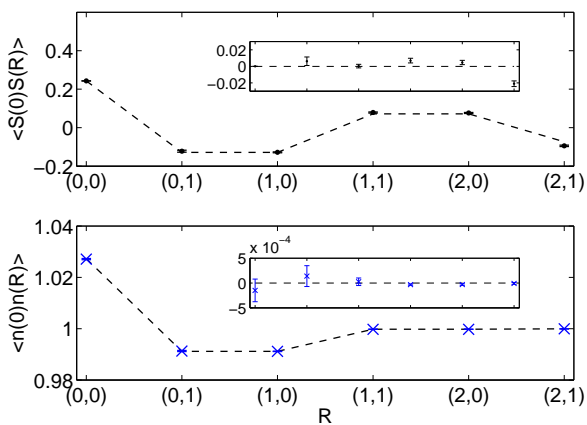} 
\caption{(Color online) Spin-spin and charge-charge correlation function of the half filled 4x2 Hubbard ladder for U/t=16 after projection. The deviation from the DMRG result (dashed line) is shown in the inset. A small systematic deviation is still present in the spin-spin correlations at large distance.}
\label{per4_16SP.fig}
\end{figure}

\begin{figure}[h]
\includegraphics[width=0.475 \textwidth]{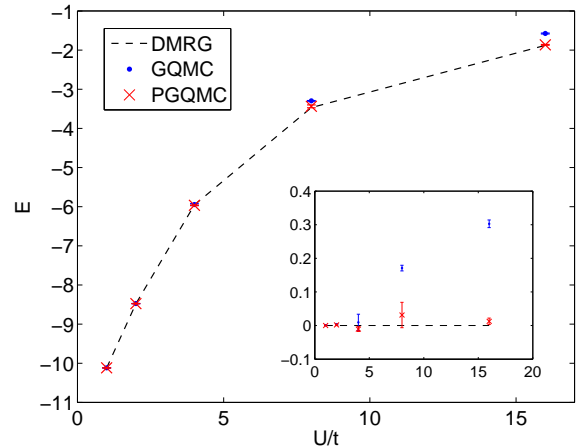}
\caption{(Color online) Energy of the half filled 4x2 Hubbard ladder in dependence of $U/t$. The deviation from the DMRG result (dashed line) is shown in the inset. Symmetry projection corrects the systematic deviations in the energy from the GQMC simulation.}
\label{Uenergy.fig}
\end{figure}

For $U/t=2$ we also obtain correct results without projection. Systematic deviations of the order $5\%$ appear for $U/t=4$, which are corrected by symmetry projection. 
As expected we observe an increase of the systematic deviations with increasing $U/t$. For $U/t=16$ symmetry projection yields the correct energy but fails to reproduce all the spin-spin correlations at large distance correctly (see Fig.~\ref{per4_16SP.fig}). In this case the average overlap of $28\%$ is rather small. Thus symmetry projection yields better results for intermediate $U/t$ but for large $U/t \ge8$ a small systematic error is still present. 
In Fig.~\ref{Uenergy.fig} we have plotted the dependence of the energy on $U/t$. Without projection the systematic error grows with increasing $U/t$, whereas the results after projection agree with the exact result for all $U/t$.

\subsubsection{$U/t=4$ and varying $L$}
\begin{figure}[b]
\includegraphics[width=0.475 \textwidth]{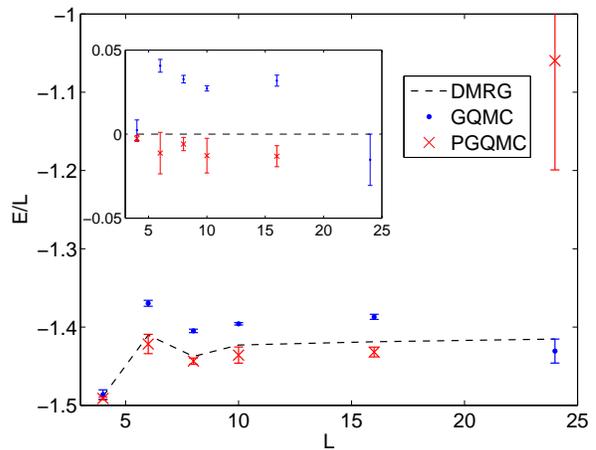}
\caption{(Color online) Energy density of the half filled two leg Hubbard ladder with $U/t=4$ in dependence of $L$. The deviation from the DMRG result (dashed line) is shown in the inset. The energy from GQMC is systematically too high (dots). The results are correct after symmetry projection for $L \le16$.  }
\label{Lenergy.fig}
\end{figure}

In this set of simulations we fixed the interaction strength to an intermediate value $U/t=4$ and varied the system length $L$. For $L>4$ we observe that the energy from the GQMC simulation is systematically too high (Fig.~\ref{Lenergy.fig}). The deviations are of order $2\%$. After symmetry projection the results are within $2\sigma$ for system sizes up to $L=16$. 

\begin{figure}[t]
\includegraphics[width=0.475 \textwidth]{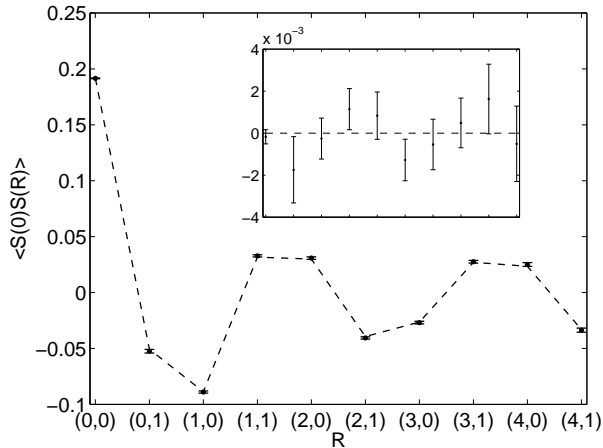} 
\caption{Spin-spin correlation function of the half filled 8x2 Hubbard model with $U/t=4$. The PGQMC (dots) results agree with the DMRG result (dashed line). The deviation from the DMRG result is shown in the inset.}
\label{SS8_4SP.fig}
\end{figure}

\begin{figure}[b]
\includegraphics[width=0.475 \textwidth]{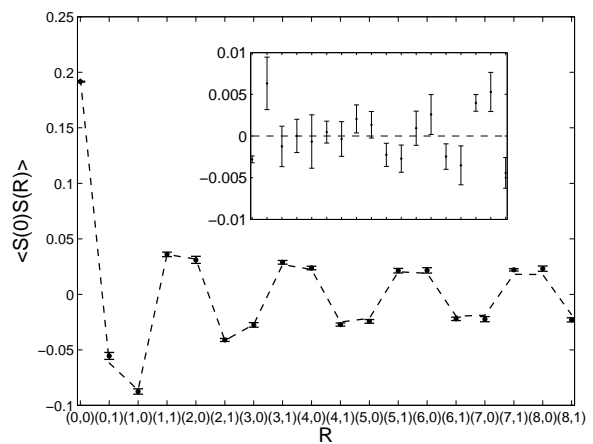} 
\caption{Spin-spin correlation function of the half filled 16x2 Hubbard model with $U/t=4$. The PGQMC (dots) results agree with the DMRG result (dashed line) for almost all distances. The deviation from the DMRG result is shown in the inset.}
\label{SS16_4SP.fig}
\end{figure}

Excellent results for the correlation functions are obtained with the PGQMC method for $L=8$ (Fig.~\ref{SS8_4SP.fig}) and $L=10$. Without projection the results are qualitatively good, systematic deviations are of order $10\%$ for the spin-spin correlations and of order $0.2\%$ for the charge-charge correlations. 
For $L=16$ (Fig. \ref{SS16_4SP.fig}) the values of the spin-spin correlations at large distances tend to be too large (in absolute value), such that systematic errors may still be present. The problem is that the overlap with the ground state sector decreases with increasing L.  For $L=4$ it is typically of order $70\%$ whereas for $L=10$ we find an overlap around $20\%$. For $L\ge16$ it is only a few percent, such that the results from the projection method are not reliable anymore and we find systematic deviations even after projecting.

At half filling the total number of particles stays constant, but the number of particles with spin up $n_\uparrow$ is not necessarily equal to the number of particles with spin down $n_\downarrow$. For big system sizes the simulation can get stuck in a configuration where $n_\uparrow \neq n_\downarrow$. In this case $S_z^{total}$ is not equal to zero and therefore the overlap with the sector $S^{total}=0$ almost vanishes. A solution to this problem is to use the quantization axis along the x-direction, leading to identical SDEs for $n_\uparrow$ and $n_\downarrow$ and therefore $n_\uparrow = n_\downarrow$ is always guaranteed. However, even with this variant the results are not satisfying. The GQMC result shows big systematic deviations in the correlation functions at large distance. The overlap even becomes negative (and small in absolute value) for some projections, because many of the projected weights are negative. This leads to cancellation between positive and negative weights, reminiscent of the sign problem in conventional QMC.

\subsubsection{Doped examples}
\begin{figure}[b]
\includegraphics[width=0.475 \textwidth]{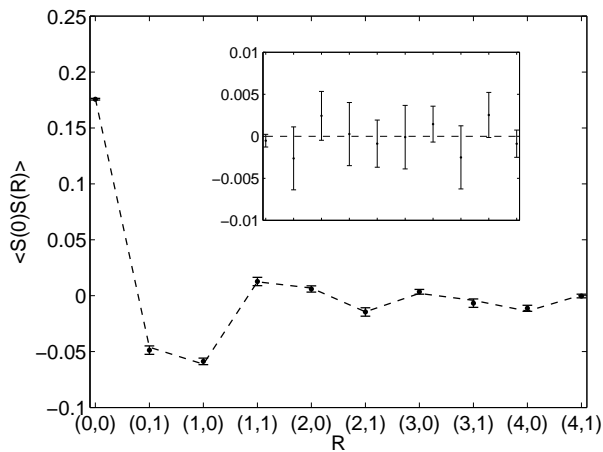} 
\caption{Spin-spin correlation function of the doped 8x2 Hubbard model with $U/t=4$ and $n_{tot}=14$. The PGQMC (dots) results agree with the DMRG result (dashed line).  The deviation from the DMRG result is shown in the inset.}
\label{SS8_n14.fig}
\end{figure} 
Next we present results for doped Hubbard ladders. The chemical potential $\mu$ is adjusted to obtain the desired number of electrons and we fix $U/t=4$.  The results for the 8x2 system with $n_{tot}=14$ are compatible with the DMRG results (Fig.~\ref{SS8_n14.fig}). However, the error bars are bigger compared to the examples at half filling. The overlap with the ground state sector is only $\approx 12\%$.

We also find agreement for the 10x2 system with $n_{tot}=18$, even if the average overlap is only $6\%$. The energy from DMRG $-16.6393$ lies within $2\sigma$ of the PGQMC result $-16.2244 \pm  0.3206$.
Slightly doped Hubbard models are known to exhibit a strong sign problem in conventional QMC, therefore it is a considerable success to obtain correct results for this case.  

\subsection{Three leg ladders}
\begin{figure}[h]
\includegraphics[width=0.475 \textwidth]{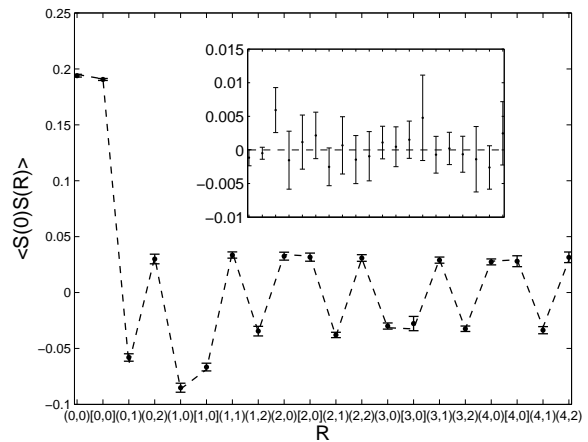} 
\caption{Spin-spin correlation function for the half filled 8x3 Hubbard model with $U/t=4$ after projection. The distances indicated by square brackets refer to distances on the middle leg of the ladder. The deviation from the DMRG result is shown in the inset.}
\label{SS3leg8_SP.fig}
\end{figure}

Three leg Hubbard ladders are critical, thus the energy gap vanishes in the thermodynamic limit. As the low lying excitations lie closer to the ground state we expect that the density matrix from the simulation will contain more admixtures of excited states than for the two leg ladders. This would result in a smaller overlap with the ground state sector and thus a less efficient symmetry projection.
We tested system sizes from $L=4$ to $L=16$.  We obtain correct results up to $L=12$ (Fig.~\ref{SS3leg8_SP.fig}).  For $L=16$ the overlap becomes too small, leading to very large error bars (see Table \ref{3leg.dat}). Note that in this case the energy obtained from GQMC alone is actually better than the energy afer the projection. However, the GQMC spin-spin correlations show large systematic deviations for large distances. Thus, neither of the two methods yield useful results for $L=16$.
 \\

\begin{table}[h]
\begin{tabular}{|c|c|c|c|c|}
\hline
L & GQMC & PGQMC & DMRG & Overlap\\
\hline
$4$    & $-9.042 \pm  0.007$& $-9.213 \pm 0.015$&  $-9.2053$ & $57\%$ \\ 
$6$ & $ -13.54 \pm  0.05$ &  $-13.69 \pm  0.10$ & $-13.7901$ & $54\%$ \\
 $8$ &  $-17.98 \pm 0.02$ & $-17.98 \pm 0.23$  & $-18.2005$ & $43\% $ \\
 $12$ & $-26.54 \pm  0.04$ &  $-27.24 \pm  0.08$  &$-27.2717$ & $21\%$\\
 $16$ & $-36.57 \pm  0.21$ & $-32.02  \pm 2.93 $ & $-36.2408$ & $2\%$\\
\hline
\end{tabular}
\caption{Energies for the Lx3 Hubbard ladder at half filling. The results from PGQMC are within $2\sigma$ up to $L=12$. 
}
\label{3leg.dat}
\end{table}

\subsection{Summary of the Hubbard ladder results}
Let us summarize the results for the Hubbard ladders:
\begin{itemize}
\item
For small system sizes GQMC yields correct results for weak interaction ($U/t\le2$). Systematic deviations for intermediate interaction strength can be fixed by symmetry projection, but only for $U/t$ not too large ($U/t\le8$). For strong interaction we find systematic errors even for the symmetry projected result.
\item
The overlap has to be big enough in order to get meaningful results. In our simulations overlaps of $> 30\%$ lead to correct results. Overlaps below $10\%$ are definitely too small. We obtained some nice results with overlaps in between $10-30\%$, but the reliability is not guaranteed.
\item 
The overlap between the GQMC density matrix and the symmetry sector of the ground state decreases with increasing system size. In our examples the overlap becomes too small for $L \ge 16$. Using more walkers may help to increase the upper limit of $L$ for the which PGQMC produces good values. 
\end{itemize}

\section{Summary and outlook}
\label{outlook}
The discussion in section \ref{systerr} suggests that the systematic errors found in the Hubbard model close to half filling stem from non-vanishing boundary terms from the partial integration step in the derivation of the SDEs. This problem has also been reported for bosonic systems \cite{Gilchrist97}, and we observe similar side effects of the boundary terms, such as spiking trajectories.  Thanks to the overcompleteness of the Gaussian operator basis it is possible to modify the SDEs without changing the physical system. The hope is to find appropriate gauges to obtain a faster decaying distribution function, such that the boundary terms vanish. 
The study of the Metropolis algorithm showed that it leads to the same systematic errors as the reconfiguration scheme of walkers.

It is important to point out that for a large parameter range of the Hubbard model GQMC yields the correct results. Therefore it would be worthwhile applying the method also to other models. Checking for spikes and slow decaying probability distributions provides an important test of the reliability of the results.

We observed that the main effect of the boundary terms is that the solution from the simulation does not exhibit all the symmetries of the Hamiltonian. By projecting the density matrix onto the ground state symmetry sector it is possible to extract the correct ground state. In order to find the limits of the symmetry projection method we have systematically tested it for Hubbard ladders and compared the results with DMRG calculations. The overlap between the density matrix and the ground state sector has to be big enough ($>30\%$) in order to obtain good results. The results agree well for systems up to 32 sites and an on site repulsion $U<10$, beyond these values the overlap becomes too small. However, we were able to obtain the correct values for doped ladders, which suffer from the negative sign problem in auxiliary field QMC. 

To conclude, even though GQMC is sign-free there are still unresolved problems. A future study will show how the boundary terms depend on the choice of gauges and if it is possible to avoid systematic errors even without projection.

Recently Aimi and Imada \cite{Aimi07} presented a new variant of the projection method, called the pre-projection method, which allows to treat bigger system sizes. Instead of projecting the density matrix after the simulation they incorporate the projection into the sampling, which leads to a better convergence towards the ground state. The price however is the occurence of negative weights, or in other words a sign problem which in many cases seems to be tractable. The results for Hubbard models up to system sizes 10x10 look very promising. For intermediate U the projected distributions decay much faster, so that no boundary terms seem to be present. As the method enables to simulate doped and frustrated systems, it is one of the most promising ground state methods for fermions currently available. However, further tests are needed to check the reliability of the method.

\begin{acknowledgments}
We acknowledge useful discussions with J. Corney, P. Drummond, M. Imada, W. P. Petersen, D. W\"urtz, E. Gull, L. Pollet. The GQMC calculations were done on the Hreidar cluster of ETH Zurich.
AK and US acknowledge support by the DFG. FFA acknowledges financial support of the DFG under the  grant number AS 120/4-2.
\end{acknowledgments}

\appendix
\section{The SDEs for the Hubbard model used in the simulation}
\label{app1}
For the derivation of the real valued, positive weighted SDEs the "fermi gauge" \cite{Corney05}  $\hat n_{ii\sigma}^2-\hat n_{ii\sigma}=0$ was used to rewrite the interaction term as 
\begin{equation}
-\frac{|U|}{2} \sum_i (\hat n_{ii\uparrow} - s \hat n_{ii\downarrow})^2 + \frac{|U|}{2} (\hat n_{ii\uparrow} + s \hat n_{ii\downarrow}), \quad s=\text{sign}(U). 
 \end{equation}
The Stratonovich stochastic differential equations for the Hubbard model read
\begin{eqnarray}
\label{sdes.eqn} d\Omega(\tau) &=& - \Omega h d\tau\\
d n_{uv\rho} &=&  A_{uv\rho} {d\tau} + \sum_{i} \left( B_{uv\rho}^{i} dW_{i} +  C_{uv\rho}^{i} dW'_{i} \right). 
\end{eqnarray}
with
\begin{eqnarray}
h &=& -t\sum_{\langle i,j \rangle \sigma} n_{ij\sigma} + U \sum_i 
n_{ii\uparrow} n_{ii\downarrow} - \mu \sum_{i\sigma} n_{ii\sigma} \nonumber \\
A_{uv\rho}&=&\frac{1}{2} \sum_{ij} (n_{uj\rho} \bar n_{iv\rho} +\bar n_{uj\rho} n_{iv\rho}) \nonumber \\
&\times& \left( t \delta_{\langle i,j \rangle} +|U|(n_{ii\rho} - sn_{ii
    -\rho}- \frac{1}{2}) \delta_{ij}+\mu\delta_{ij} \right)  \nonumber \\
B_{uv\rho}^i&=&\sqrt{\frac{|U|}{2}}(n_{ui\uparrow}\bar
n_{iv\uparrow}\delta_{\rho\uparrow} -s n_{ui\downarrow}\bar
n_{iv\downarrow}\delta_{\rho\downarrow})\nonumber \\
C_{uv\rho}^i&=&\sqrt{\frac{|U|}{2}}(\bar n_{ui\uparrow}
n_{iv\uparrow}\delta_{\rho\uparrow} -s \bar n_{ui\downarrow}
n_{iv\downarrow}\delta_{\rho\downarrow}),
\end{eqnarray}
where ${\langle i,j \rangle}$ denotes nearest neighbor pairs and the noise terms $dW_i$ are defined by the correlations $\langle
dW_i dW_{i'}\rangle = d\tau \delta_{i i'} $
and the mean $\langle dW_i(\tau)\rangle=0$. We use the notation $\bar n_{ui\sigma}= \delta_{ui} - n_{ui\sigma}$.

The drift term in the Ito formulation reads
\begin{eqnarray}
A_{uv\rho}^{Ito}&=&\frac{1}{2} \sum_{ij} (n_{uj\rho} \bar n_{iv\rho} +\bar n_{uj\rho} n_{iv\rho}) \nonumber \\
&\times& \left( t \delta_{\langle i,j \rangle} -  U n_{ii-\rho}\delta_{ij}+\mu\delta_{ij}\right).
\end{eqnarray}

\section{Adaptive time step}
\label{ats}
To reduce the error from the time discretization of the SDEs we use an adaptive time step. Initially we choose a maximal step size $\Delta \tau_{\text{max}}$. Whenever any element of the drift term exceeds a certain threshold,
\begin{equation}
 \max_{uv\rho} (A_{uv\rho} \cdot \Delta \tau) > u_{\text{max}},
 \end{equation}
we divide the current time interval $\Delta \tau$ by 2 and perform 2 update steps with a reduced step size $\Delta \tau/2$. In each of these two steps the above condition is tested again and the step size is further divided by 2 if necessary. By proceeding in the same way for each reduced interval the step size can become arbitrarily small (limited by the machine precision). For example, the interval $\Delta \tau$ may be split into 4 smaller intervals with step sizes $\Delta \tau/2 + \Delta \tau/8 + \Delta \tau/8 + \Delta \tau/4$. We usually choose $u_{max}$ between $0.01$ and $0.05$.

\bibliography{SDE}

\begin{thebibliography}{17}
\expandafter\ifx\csname natexlab\endcsname\relax\def\natexlab#1{#1}\fi
\expandafter\ifx\csname bibnamefont\endcsname\relax
  \def\bibnamefont#1{#1}\fi
\expandafter\ifx\csname bibfnamefont\endcsname\relax
  \def\bibfnamefont#1{#1}\fi
\expandafter\ifx\csname citenamefont\endcsname\relax
  \def\citenamefont#1{#1}\fi
\expandafter\ifx\csname url\endcsname\relax
  \def\url#1{\texttt{#1}}\fi
\expandafter\ifx\csname urlprefix\endcsname\relax\def\urlprefix{URL }\fi
\providecommand{\bibinfo}[2]{#2}
\providecommand{\eprint}[2][]{\url{#2}}

\bibitem[{\citenamefont{Assaad et~al.}(2005)\citenamefont{Assaad, Werner,
  Corboz, Gull, and Troyer}}]{Assaad05}
\bibinfo{author}{\bibfnamefont{F.~F.} \bibnamefont{Assaad}},
  \bibinfo{author}{\bibfnamefont{P.}~\bibnamefont{Werner}},
  \bibinfo{author}{\bibfnamefont{P.}~\bibnamefont{Corboz}},
  \bibinfo{author}{\bibfnamefont{E.}~\bibnamefont{Gull}}, \bibnamefont{and}
  \bibinfo{author}{\bibfnamefont{M.}~\bibnamefont{Troyer}},
  \bibinfo{journal}{Phys. Rev. B} \textbf{\bibinfo{volume}{72}},
  \bibinfo{pages}{224518} (\bibinfo{year}{2005}).

\bibitem[{\citenamefont{Troyer and Wiese}(2005)}]{Troyer05}
\bibinfo{author}{\bibfnamefont{M.}~\bibnamefont{Troyer}} \bibnamefont{and}
  \bibinfo{author}{\bibfnamefont{U.-J.} \bibnamefont{Wiese}},
  \bibinfo{journal}{Phys. Rev. Lett.} \textbf{\bibinfo{volume}{94}},
  \bibinfo{pages}{170201} (\bibinfo{year}{2005}).

\bibitem[{\citenamefont{Corney and Drummond}(2004)}]{Corney04}
\bibinfo{author}{\bibfnamefont{J.~F.} \bibnamefont{Corney}} \bibnamefont{and}
  \bibinfo{author}{\bibfnamefont{P.~D.} \bibnamefont{Drummond}},
  \bibinfo{journal}{Phys. Rev. Lett.} \textbf{\bibinfo{volume}{93}},
  \bibinfo{pages}{260401} (\bibinfo{year}{2004}).

\bibitem[{\citenamefont{Corney and Drummond}(2006)}]{corney_large}
\bibinfo{author}{\bibfnamefont{J.~F.} \bibnamefont{Corney}} \bibnamefont{and}
  \bibinfo{author}{\bibfnamefont{P.~D.} \bibnamefont{Drummond}},
  \bibinfo{journal}{Phys. Rev. B} \textbf{\bibinfo{volume}{73}},
  \bibinfo{pages}{125112} (\bibinfo{year}{2006}).

\bibitem[{\citenamefont{Assaad et~al.}(2006)\citenamefont{Assaad, Corboz, Gull,
  Petersen, Troyer, and Werner}}]{Assaad06}
\bibinfo{author}{\bibfnamefont{F.~F.} \bibnamefont{Assaad}},
  \bibinfo{author}{\bibfnamefont{P.}~\bibnamefont{Corboz}},
  \bibinfo{author}{\bibfnamefont{E.}~\bibnamefont{Gull}},
  \bibinfo{author}{\bibfnamefont{W.~P.} \bibnamefont{Petersen}},
  \bibinfo{author}{\bibfnamefont{M.}~\bibnamefont{Troyer}}, \bibnamefont{and}
  \bibinfo{author}{\bibfnamefont{P.}~\bibnamefont{Werner}}, in
  \emph{\bibinfo{booktitle}{Effective models for low-dimensional strongly
  correlated systems}} (\bibinfo{year}{2006}), vol. \bibinfo{volume}{816} of
  \emph{\bibinfo{series}{AIP Conference Proceedings}}, pp.
  \bibinfo{pages}{204--231}.

\bibitem[{\citenamefont{Gilchrist et~al.}(1997)\citenamefont{Gilchrist,
  Gardiner, and Drummond}}]{Gilchrist97}
\bibinfo{author}{\bibfnamefont{A.}~\bibnamefont{Gilchrist}},
  \bibinfo{author}{\bibfnamefont{C.~W.} \bibnamefont{Gardiner}},
  \bibnamefont{and} \bibinfo{author}{\bibfnamefont{P.~D.}
  \bibnamefont{Drummond}}, \bibinfo{journal}{Phys. Rev. A}
  \textbf{\bibinfo{volume}{55}}, \bibinfo{pages}{3014} (\bibinfo{year}{1997}).

\bibitem[{\citenamefont{Deuar and Drummond}(2002)}]{Deuar02}
\bibinfo{author}{\bibfnamefont{P.}~\bibnamefont{Deuar}} \bibnamefont{and}
  \bibinfo{author}{\bibfnamefont{P.~D.} \bibnamefont{Drummond}},
  \bibinfo{journal}{Phys. Rev. A} \textbf{\bibinfo{volume}{66}},
  \bibinfo{pages}{033812} (\bibinfo{year}{2002}).

\bibitem[{\citenamefont{Buonaura and Sorella}(1998)}]{sorella}
\bibinfo{author}{\bibfnamefont{M.~C.} \bibnamefont{Buonaura}} \bibnamefont{and}
  \bibinfo{author}{\bibfnamefont{S.}~\bibnamefont{Sorella}},
  \bibinfo{journal}{Phys. Rev. B} \textbf{\bibinfo{volume}{57}},
  \bibinfo{pages}{11446} (\bibinfo{year}{1998}).

\bibitem[{\citenamefont{Schollw{\"o}ck}(2005)}]{Schollwoeck05}
\bibinfo{author}{\bibfnamefont{U.}~\bibnamefont{Schollw{\"o}ck}},
  \bibinfo{journal}{Rev. Mod. Phys.} \textbf{\bibinfo{volume}{77}},
  \bibinfo{pages}{259} (\bibinfo{year}{2005}).

\bibitem[{\citenamefont{White}(1992)}]{White92}
\bibinfo{author}{\bibfnamefont{S.~R.} \bibnamefont{White}},
  \bibinfo{journal}{Phys. Rev. Lett.} \textbf{\bibinfo{volume}{69}},
  \bibinfo{pages}{2863} (\bibinfo{year}{1992}).

\bibitem[{\citenamefont{Aimi and Imada}(2007)}]{Aimi07}
\bibinfo{author}{\bibfnamefont{T.}~\bibnamefont{Aimi}} \bibnamefont{and}
  \bibinfo{author}{\bibfnamefont{M.}~\bibnamefont{Imada}},
  \bibinfo{journal}{arxiv.org/abs/0704.3792}  (\bibinfo{year}{2007}).

\bibitem[{\citenamefont{Bir{\'o} and Jakov{\'a}c}(2005)}]{Biro05}
\bibinfo{author}{\bibfnamefont{T.~S.} \bibnamefont{Bir{\'o}}} \bibnamefont{and}
  \bibinfo{author}{\bibfnamefont{A.}~\bibnamefont{Jakov{\'a}c}},
  \bibinfo{journal}{Phys. Rev. Lett.} \textbf{\bibinfo{volume}{94}},
  \bibinfo{pages}{132302} (\bibinfo{year}{2005}).

\bibitem[{\citenamefont{Levy and Solomon}(1996)}]{Levy96}
\bibinfo{author}{\bibfnamefont{M.}~\bibnamefont{Levy}} \bibnamefont{and}
  \bibinfo{author}{\bibfnamefont{S.}~\bibnamefont{Solomon}},
  \bibinfo{journal}{International Journal of Modern Physics C}
  \textbf{\bibinfo{volume}{7}}, \bibinfo{pages}{595} (\bibinfo{year}{1996}).

\bibitem[{\citenamefont{Corney and Drummond}(2005)}]{Corney05}
\bibinfo{author}{\bibfnamefont{J.~F.} \bibnamefont{Corney}} \bibnamefont{and}
  \bibinfo{author}{\bibfnamefont{P.~D.} \bibnamefont{Drummond}},
  \bibinfo{journal}{cond-mat/0411712}  (\bibinfo{year}{2005}).

\bibitem[{\citenamefont{Dowling et~al.}(2005)\citenamefont{Dowling, Davis,
  Drummond, and Corney}}]{dowling}
\bibinfo{author}{\bibfnamefont{M.~R.} \bibnamefont{Dowling}},
  \bibinfo{author}{\bibfnamefont{M.~J.} \bibnamefont{Davis}},
  \bibinfo{author}{\bibfnamefont{P.~D.} \bibnamefont{Drummond}},
  \bibnamefont{and} \bibinfo{author}{\bibfnamefont{J.~F.}
  \bibnamefont{Corney}}, \bibinfo{journal}{arXiv:quant-ph/0507003}
  (\bibinfo{year}{2005}).

\bibitem[{\citenamefont{Hastings}(1970)}]{hasting}
\bibinfo{author}{\bibfnamefont{W.~K.} \bibnamefont{Hastings}},
  \bibinfo{journal}{Biometrika} \textbf{\bibinfo{volume}{57}},
  \bibinfo{pages}{97} (\bibinfo{year}{1970}).

\bibitem[{\citenamefont{McCulloch and Gulacsi}(2002)}]{McCulloch02}
\bibinfo{author}{\bibfnamefont{I.~P.} \bibnamefont{McCulloch}}
  \bibnamefont{and} \bibinfo{author}{\bibfnamefont{M.}~\bibnamefont{Gulacsi}},
  \bibinfo{journal}{Europhys. Lett.} \textbf{\bibinfo{volume}{57}},
  \bibinfo{pages}{852} (\bibinfo{year}{2002}).

\end{thebibliography}
\end{document}